\documentclass[aps,showpacs,twocolumn,longbibliography,nofootinbib,secnumarabic,superscriptaddress]{revtex4-1}

\usepackage{graphicx}
\usepackage{dcolumn}
\usepackage{bm}
\usepackage{braket}
\usepackage{leftidx}
\usepackage{cancel}
\usepackage{amsmath}
\usepackage{color}
\usepackage[mathcal]{eucal}
\usepackage{float}

\begin{document}
\date{\today}
\title{Continuously-tunable Cherenkov-radiation-based detectors  via plasmon index control}

\author{Mehmet G\"{u}nay}
\affiliation{Department of Nanoscience and Nanotechnology, Faculty of Arts and Science, Burdur Mehmet Akif Ersoy University, 15030 Burdur, Turkey}
\affiliation{Institute of Nuclear Sciences, Hacettepe University, 06800, Ankara, Turkey}

\author{You-Lin Chuang}
\affiliation{Physics Division, National Center for Theoretical Sciences, Hsinchu 30013, Taiwan}

\author{Mehmet Emre Tasgin}\thanks{Corresponding Author: metasgin@hacettepe.edu.tr  and metasgin@gmail.com}
\affiliation{Institute of Nuclear Sciences, Hacettepe University, 06800, Ankara, Turkey}

\begin{abstract}
A recent study [PRB 100, 075427 (2019)], finally, demonstrated plasmon-analog of refractive index enhancement in metal nanostructures, which has already been studied in atomic clouds for several decades. Here, we simply utilize this phenomenon for achieving continuously-tunable enhanced Cherenkov radiation in metal nanostructures. Beyond enabling Cherenkov radiation from slow-moving particles, or increasing its intensity, the phenomenon can be used in continuous-tuning the velocity cutoff of particles contributing to the Cherenkov radiation. {More influentially,} this allows a continuously-tunable analysis of the contributing particles {as if the data is collected from many different detectors, which enables data correction}. The phenomenon can also be integrated into lattice metal nanostructures, for {continuous medium tuning}, where a high density of photonic states is present and the threshold for the Cherenkov radiation can even be lifted. {Additionally, vanishing absorption can heal radiation angle distortion effects caused by the metallic absorption.}
\end{abstract}
\maketitle

\section{Introduction}

A charged particle, moving with a constant velocity in a dielectric medium, emits the well-known Cherenkov radiation~(CR)~\cite{cherenkov1934visible,frank1991coherent} when its velocity~($v$) exceeds the phase velocity~($v_{\rm ph}=c/n(\omega)$) of light, i.e. $v>v_{\rm ph}$, in this medium~\cite{landau2013electrodynamics}. Such a condition is met for particles of energy in the order of a hundred keV, which can be generated in nuclear processes and particle accelerators ---also used for free-electron lasers~\cite{fisch1992operation,sartini2010nuclear}. CR can be utilized for particle detection purposes~\cite{ahmed2007physics,icecube2013evidence}. Velocity distribution of emitted particles can be characterized by the angle $\theta_{\rm \scriptscriptstyle CR}$ of CR via $\cos\theta_{\rm \scriptscriptstyle CR}=v_{\rm ph}/v$~\cite{landau2013electrodynamics}. Detection of slower-moving particles, however, necessitates media with much larger refractive indices.

Recent developments in the control and manufacturing of nanostructures~(NSs)~\cite{Nanomanufacturing2017,Enzo2018lightMatter,tame2013quantum} enabled particle detectors based on metal nanostructures~(MNSs)~\cite{liu2012surface,georgescu2012vcerenkov,zhao2015cherenkov}. CR near a thin metal film deposited on dielectric~\cite{liu2012surface} and periodic structures~\cite{lin2018controlling,JoannopoulosScience2013,liu2017dispersive} of metallic nanoarrays~\cite{vorobev2012nondivergent,tyukhtin2014radiation,so2010cerenkov} are very different from a medium of a uniform index. Lattices of MNSs facilitate hyperbolic metamaterials~\cite{Silveirinha2012cherenkov,silveirinha2017metamaterials}. In these metamaterials enhanced photonic density of states boosts the intensity of CR. Moreover, the threshold for CR radiation can be lifted in these metamaterials~\cite{liu2017integrated}. Such a periodic medium is experimentally demonstrated to emit CR for electron energies as lows as 0.25 keV~\cite{Silveirinha2012cherenkov,liu2017integrated}. Furthermore, left-handed metamaterials demonstrate a reversed CR~\cite{Cherenkov_negative2003,Cherenkov_negativePRL2009Experiment}. The progresses in Cherenkov radiating metamaterials have stimulated the use of CR imaging for biological and medical applications~\cite{shaffer2017utilizing}.

\begin{figure}
\begin{center}
\includegraphics[scale=0.55]{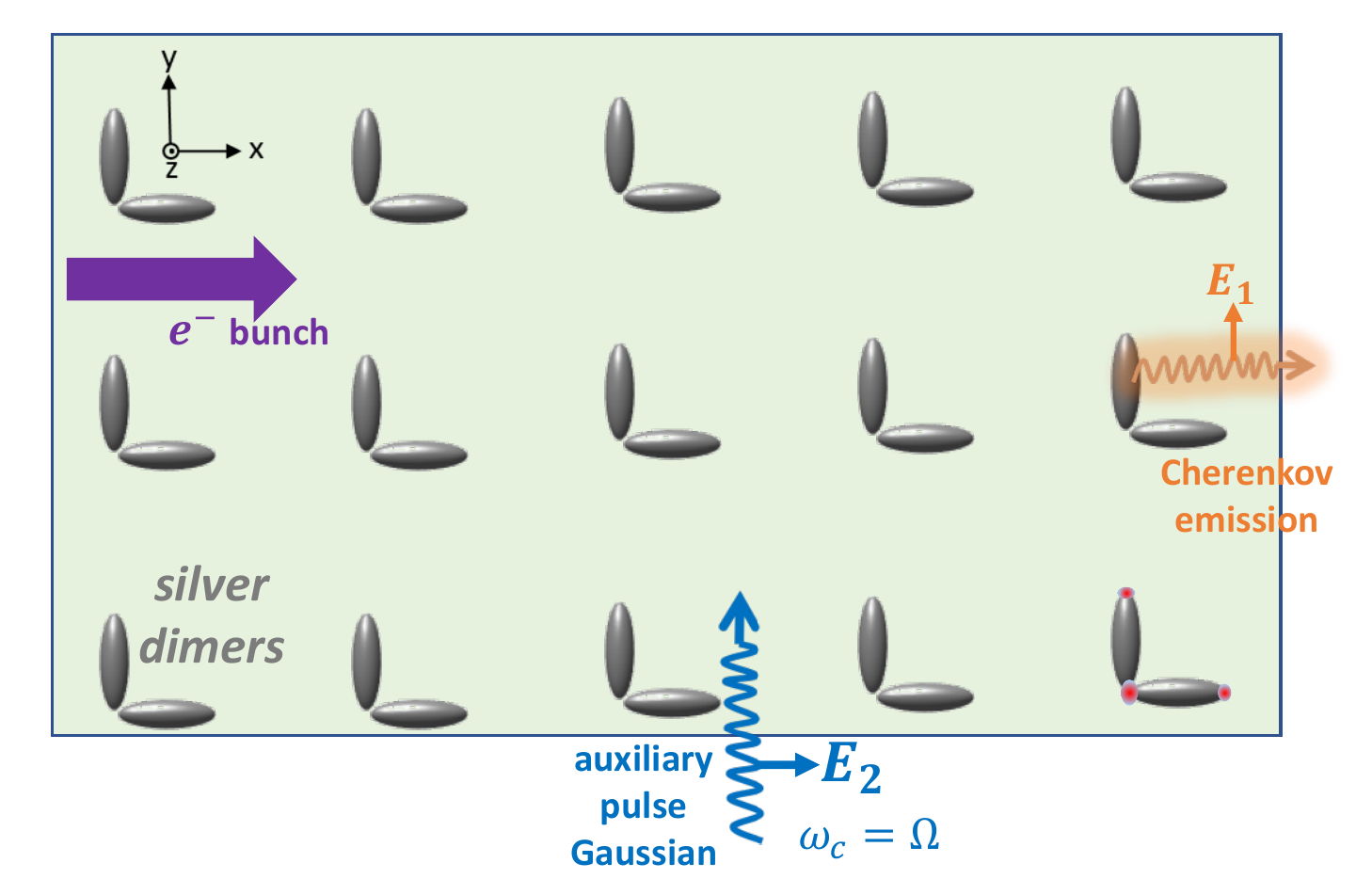} 
\caption{  Silver nanoparticle dimers, aligned in the x and y directions, are illuminated with an x-polarized auxiliary Gaussian pulse of frequency $\Omega$. The x-aligned silver nanoparticles interact with the incident field very strongly to produce nm-size hotspots and yields an enhancement in the polarization of the y-aligned nanoparticle via strong interaction between the two nanoparticles. A slow-moving charged particle traveling in the medium~\cite{yurtsever2008formation,silveirinha2017metamaterials,GinisPRL2014Anisotropic,Silveirinha2012cherenkov,morgado2015analytical} emits a y-polarized Cherenkov radiation via the index enhancement. Polarization density when the auxiliary pulse is off and on are given in Fig.~\ref{fig_w_beta_I_1}(a.i) and Fig.~\ref{fig_w_beta_I_1}(b.i), respectively. 
\label{fig_dimers}}
\end{center}
\end{figure}

Besides facilitating the movements in CR imaging, MNSs provided a medium also for observing plasmon analogs~\cite{tame2013quantum} of electromagnetically-induced transparency~(EIT) like effects~\cite{luk2010fano,limonov2017fano,peng2014and}, originally observed for 3 or more level atoms~\cite{fleischhauer1992resonantly,fleischhauer2005electromagnetically,ScullyZubairyBook}. Fano resonances~\cite{luk2010fano,limonov2017fano,panaro2014dark}, plasmon-analog of EIT, and nonlinear response enhancement~\cite{butet2014fano,singh2016enhancement,TasginFanoBook2018,postaci2018silent} have been demonstrated in MNSs. Very recently, finally, the plasmon-analog of refractive index enhancement is also demonstrated via simple analytical calculations which are supported by the exact solutions of the 3D Maxwell equations~\cite{plasmon_index_enhancement}. Lavrinenko and colleagues demonstrate~\cite{plasmon_index_enhancement} that linear response of a metal nanorod (y-aligned) to the applied electric field can be controlled, in particular be enhanced, via interacting with a second (perpendicular, x-aligned) nanorod, see dimers in Fig.~\ref{fig_dimers}. This happens when the second (x-aligned) nanorod is driven with a pump which does not couple to (not excite) the first (y-aligned) nanorod directly. The interaction between the two nanorods is provided by the hotspot~(near-field) of the second nanorod which relies at the intersection of the two nanorods.

In this paper, we discuss the utilization of this index control scheme to the plasmonic Cherenkov radiation-based detectors employing metal nanoparticles. The phenomenon~\cite{plasmon_index_enhancement} can be used to tune (e.g. to decrease) the phase velocity~\footnote{not the group velocity, unlike Refs.~\cite{carusotto2001slow,EITCherenkov}}, of the light propagation in a particle detector continuously. We show that the phenomenon can be utilized (i) to enhance the total Cherenkov radiation, (ii) to enable the Cherenkov radiation from slow-moving particles, (iii) to control the wavelength of CR, and (iv) to gain control over the cutoff velocity of the particles emitting the Cherenkov radiation. {Actually, more influential than (i)-(iv) is: continuous tuning of the detector, after it is manufactured, allows a particle velocity distribution analysis as if there exist many detectors.}

We consider an ensemble of metal nanorod dimers, see Fig.~\ref{fig_dimers}, and calculate a polarization density~$P(\omega)$ from the enhanced dipole moment responses of such dimers. The dipole moments of x-aligned nanorods are controlled by an x-polarized auxiliary~(aux) Gaussian pulse.  The x-polarized nanorod couples to the y-polarized nanorod and tunes its response to an applied (probe) E-field of polarization along the y-direction. (The y-aligned nanorod can be excited with a y-polarized field.) { Thus, an x-polarized aux beam ($E_2$) can continuously tune the refractive index of the medium for a y-polarized CR emission. The charged particles travel through the medium in the x-direction~\cite{yurtsever2008formation,silveirinha2017metamaterials,GinisPRL2014Anisotropic,Silveirinha2012cherenkov,morgado2015analytical}. Enhanced y-polarized CR emission and its angle, measured in the x-z plane, can be recorded using a y-polarization selector (filter). We discuss the phenomenon in the following sections in details.}

{
The main input of the presented scheme is neither the enhancement of the Cherenkov radiation intensity nor bringing the velocity threshold to slower-moving particles. In current particle detectors, employing metal nanostructures, CR intesity is already enhanced due to the increased density of states
~\cite{Silveirinha2012cherenkov} and the velocity threshold can even be lifted~\cite{silveirinha2017metamaterials,liu2017integrated}. None of these current detectors, however, can provide a continuous tuning, thus a continuous spectral analysis, of the particles' velocities via CR. 

We can make the functionality of the new scheme more visible as follows. A conventional (CR-based) particle detector, whether employing the nanoparticles or not, takes (records) the emission spectral data for the \textit{fixed parameters} of the manufactured detector; even if the CR spectrum is enhanced and the threshold is lifted. Then, it analyses the particle velocity distribution etc. The particle detector, we study in this manuscript, does the following. It can record data on the velocity distribution of particles as if one collects data on the velocity distribution of the source from many particle detectors, in principle infinite, each having different fixed parameters (e.g. indices). This way, comparing all such data sets, e.g., one can make corrections on the errors of data~\cite{DataCorrectionDetector} occurring due to non-uniform absorption, or other unpredictable issues. Such a game changing continuously-tunable tool/device, to our best knowledge, is not present in the new-generation CR-based particle detectors.
}

{ The index control mechanism we employ here provides one another significant advantage in CR-based particle velocity detection. In conventional CR-based detectors, employing periodic metal nanostructures, there exists a serious amount of loss. This severely distorts (e.g. broaden) the relation between the Cherenkov angle and the particle velocity on which the particle detection is based~\cite{GinisPRL2014Anisotropic,lin2018NaturePhys}. In the index control scheme, we study here, however, the enhanced index can be achieved at vanishing/reduced absorption, see Fig.\ref{fig_Lavrinenko}. Vanishing/reduced absorption at different parameters can be achieved by using aux pulses operating at different frequencies.

}

%
%
%
%

We also present the contribution to the power of CR from different particle velocities of a ${}^{18}F$ emission~\cite{levin1999calculation} for different enhancements of the refractive index. We further calculate the angle $\theta_{\rm \scriptscriptstyle CR}$ for particle moving at different velocities.

{ Here, we consider a { periodic} spatial distribution of the nanorod dimer centers. All dimers are aligned along a given direction. Such kind of periodic structures  can be achieved by e-beam lithography techniques~\cite{ebeam_high_speed_2016}. Although here we consider a periodic ordering of the dimers, arbitrary distributions of such nanodimers can be achieved also in solutions when one of the nanorods of the dimer is manufactured via a magnetized metallic material~\cite{anderson2019magnetic,Enzo2017giantMagnetic}. This can be useful for in vitro CR imaging~\cite{shaffer2017utilizing}. Here we do not consider the extra enhancement effects possible to appear due to periodicity, in this first demonstration of such a game-changing device. Nevertheless, periodicity of the dimers is expected to increase the CR emission via enhanced density of photonic states, a feature already appearing in their no-index-controlled counterparts~\cite{fleischhauer1992resonantly,fleischhauer2005electromagnetically,ScullyZubairyBook}. } Using lattice structures of index controlled dimers can provide not only an enhanced emission, compared to the metal nanoarrays~\cite{liu2017integrated}, without a CR threshold, { but also make their CR emission features continuously tunable~\footnote{ Study of such periodic structures with the index-controlled scheme is also possible to lead very interesting and naive features, such as continuous tuning of band diagrams and possible jumps of such features at certain index values.}.   } Although we consider the basic dimer structure Ref.~\cite{plasmon_index_enhancement} studies originally; nanostructures, easier to manufacture, can display similar features with such dimers.

It is important to note that: in this work, we aim merely to provide a basic and the \textit{first}, ``proof of principle", demonstration of an important utilization of plasmon index-enhancement (control) scheme~\cite{Silveirinha2012cherenkov} for Cherenkov radiation applications. { We present the validity of the model we use in Appendix~\ref{sec:Appendix-validity}, which is already demonstrated for nanostructures of similar sizes~\cite{yurtsever2008formation}. There may, certainly, appear some small deviations from the mean polarization method. But, here we aim to bring a new and influential mechanism into light with basic and valid arguments~App.~\ref{sec:Appendix-validity}.}

{ The paper is organized as follows. In Sec.~\ref{sec:index_enhancement}, we describe how the (auxiliary) x-polarized pulse can control the refractive index of the medium for an x-polarized light propagation. We review the results of Ref.~\cite{Silveirinha2012cherenkov}, where a single frequency is considered, and generalize them to the case a Gaussian control pulse is employed. In Sec.~\ref{sec:CR}, we present the utilization of the refractive index control to the continuous spectral analysis of a radioactive source emitting charged particles. Sec.~\ref{sec:summary} contains our summary.}

\section{Index Enhancement} \label{sec:index_enhancement}

\begin{figure}
\begin{center}
\includegraphics[scale=0.61]{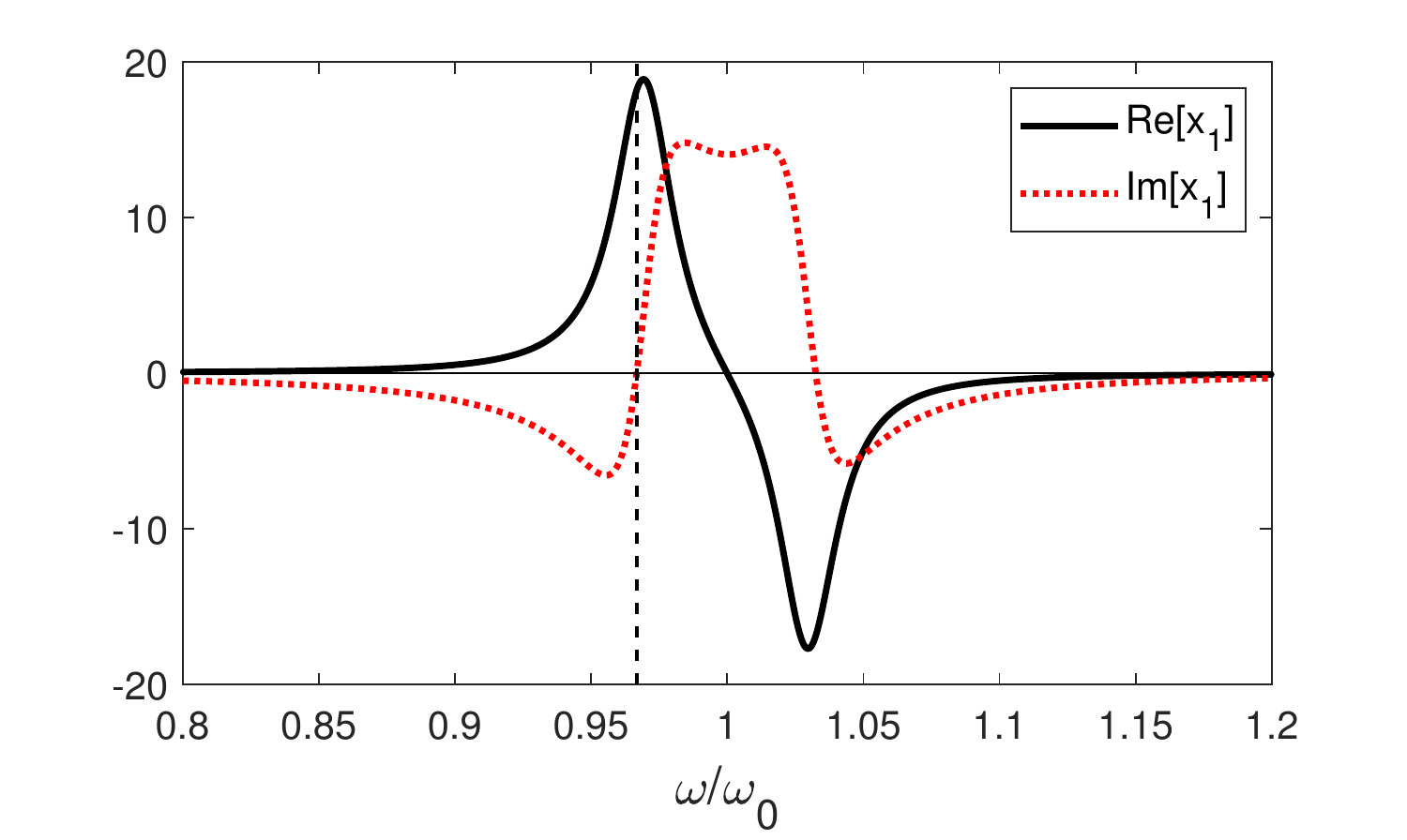} 
\caption{Enhanced oscillation strength~\cite{plasmon_index_enhancement} of the y-aligned nanorod in the presence of an auxiliary pulse $E_2$, oscillating at the same frequency with the probe pulse $E_1(t)=E_1e^{-i\omega t}$. Without the enhancement scheme, maximum $x_1$ attains a value of 0.2 only. {  At $\Omega=0.967\omega_0$ the polarization enhanced with vanishing absorption.} { The phase of the aux pump is fixed to $\phi=\pi/2$, in Eq.~(\ref{x1_steady}), as in Ref.~\cite{plasmon_index_enhancement} . } \label{fig_Lavrinenko}}
\end{center}
\end{figure}

Lavrinenko and colleagues show~\cite{plasmon_index_enhancement} that response of a y-aligned metal nanorod, to a field frequency $\omega$, can be enhanced when it is coupled to a perpendicular (x-aligned) nanorod which is driven by an  x-polarized auxiliary pulse of the same frequency $\omega$. Owing to the structure of the localized surface plasmon resonances, the x-polarized aux pulse couples (most efficiently) merely to the x-aligned nanorod, and not to the y-aligned nanorod. The y-polarized nanorod, however, couples to the plasmon mode of the x-aligned nanorod at the hotspot which appears at the intersection of the two nanorods. Ref.~\cite{plasmon_index_enhancement} shows that (see Fig.~\ref{fig_Lavrinenko}) the polarization of the y-aligned nanorod ---which  can be driven only with a y-polarized light--- responds to the incident field $E_1$ with an enhanced polarization about a given frequency { $\Omega=0.967\omega_0$}. $\omega_0$ is the resonance of the nanorods. At $\omega=\Omega$, the polarization is enhanced with a vanishing absorption, a phenomenon also observed in atomic ensembles using aux microwave~\cite{fleischhauer1992resonantly,ScullyZubairyBook} or optical~\cite{yavuz2005refractive,proite2008refractive} pulses. Although Ref.~\cite{plasmon_index_enhancement} demonstrates the phenomenon for the coupling of two (identical) nanorods, as becomes apparent below, the phenomenon can be observed for the coupling of different plasmon resonances with other configurations.

The system of two coupled nanorods is treated by two coupled oscillators, as, this model is sufficient to explain almost all fundamental plasmon interaction behaviors, e.g. in Fano resonances~\cite{Pelton2010OptExp,lovera2013mechanisms}, except second-quantized features~\cite{finazzi2012plasmon}. The results are supported also with the exact solution of 3D Maxwell equations, i.e. FDTD simulations.

The polarizations (plasmon oscillations) of the plasmon modes of the two perpendicular nanorods can be described with two coupled oscillators
\begin{eqnarray}
\ddot{x}_1+\gamma_1 \dot{x}_1 + \omega_1^2 x_1 - gx_2 = \tilde{F}_1(t), \label{x1}
\\
\ddot{x}_2+\gamma_2 \dot{x}_2 + \omega_2^2 x_1 - gx_1 = \tilde{F}_2(t), \label{x2}
\end{eqnarray}
where $\gamma_1=\gamma_2=0.026\omega_0$, $\omega_1=\omega_2=\omega_0$ and $g=0.06\omega_0$~\cite{plasmon_index_enhancement} are the damping rates, resonances and the coupling between the two plasmon oscillations of the two nanorod, respectively~{ \footnote{  While Ref.~\cite{plasmon_index_enhancement} uses the values $\gamma_1=\gamma_2=0.05\omega_0$ for the damping rates, here we use the scaled linewidth of silver obtained from the experimental data} }. $\omega_0$ is the resonance of both nanorods. The second nanorod is driven by the field $\tilde{F}_2(t)=F_2e^{-i\phi}e^{-i\omega t}$ and the response of the $x_1/F_1$ is investigated, in Ref.~\cite{plasmon_index_enhancement}, when a y-polarized electric field $\tilde{F}_1(t)=F_1e^{-i\omega t}$ is applied on the first nanorod oscillating with the same frequency $\omega$. In such a case an analytical solution for the plasmon oscillation $x_1(t)=x_1 e^{-i\omega t}$
\begin{equation}
x_1=\frac{\delta_2F_1+ge^{-i\phi}F_2}{\delta_1\delta_2-g^2}
\label{x1_steady}
\end{equation}
can be obtained with $\delta_i=\omega_i^2-\omega^2-i\gamma_i\omega$. $x_1$ is plotted in Fig.~\ref{fig_Lavrinenko}~\cite{plasmon_index_enhancement} for $F_2/\omega_0^2=1$, $F_1/\omega_0^2=0.01$ and $\phi=\pi/2$. Polarization density, (dipole moment)/volume~\cite{griffiths2005introduction}, of the medium to a y-polarized field is given by
\begin{equation}
\chi(\omega)=f\omega_0^2\frac{\delta_2+e^{-i\phi}E_2/E_1}{\delta_1\delta_2-g^2},
\label{chi}
\end{equation}
where $f$ is the dimensionless oscillator strength determined by the density of the dimers. We calculate $f$ as follows. The polarizibility of a nanorod (ellipsoid), in units of volume, can be calculated analytically~\cite{bohren2008absorption,ngom2009novel} as $\alpha(\omega)=\nu \: (\epsilon(\omega)-\epsilon_h)/[ \epsilon_h + r(\epsilon(\omega)-\epsilon_h) ]$ with $r=(1-e^2)/e^2\{-1+ln[(1+e)/(1-e)]/2e \}$. Here, $\nu$ is the volume of the nanorod.

We use $L$=30 nm for the length and $b$=10 nm the width of the ellipsoid as considered in Ref.~\cite{plasmon_index_enhancement}. Polarization density for applied E-field (without enhancement) can be obtained via $\chi(\omega)=P/E=\rho \:\alpha$, where $\rho$ is the number density of such nanorods, which we set $\rho\: \nu$=0.02. Here, we also use the experimental dielectric function $\epsilon(\omega)$ of silver~\cite{plasmon_index_enhancement,hohenester2012mnpbem} ($\gamma_1=\gamma_2=0.026\omega_0$) and obtain $f$=0.23 in Eq.~(\ref{chi}) by setting $\chi(\omega=\omega_0,g=0)=f/(\gamma_1/\omega_0)$ equal to $\chi(\omega_0)=\rho\: \alpha(\omega_0)$. Actually, for a ``proof of principle" demonstration of the benefits of the index-enhancement scheme for particle detectors, or CR imaging, such a tidy choice for $\epsilon(\omega)$ is not necessary.

\begin{figure}
\begin{center}
\includegraphics[trim=1.8cm 0 1cm 0 ,clip,width=0.5\textwidth]{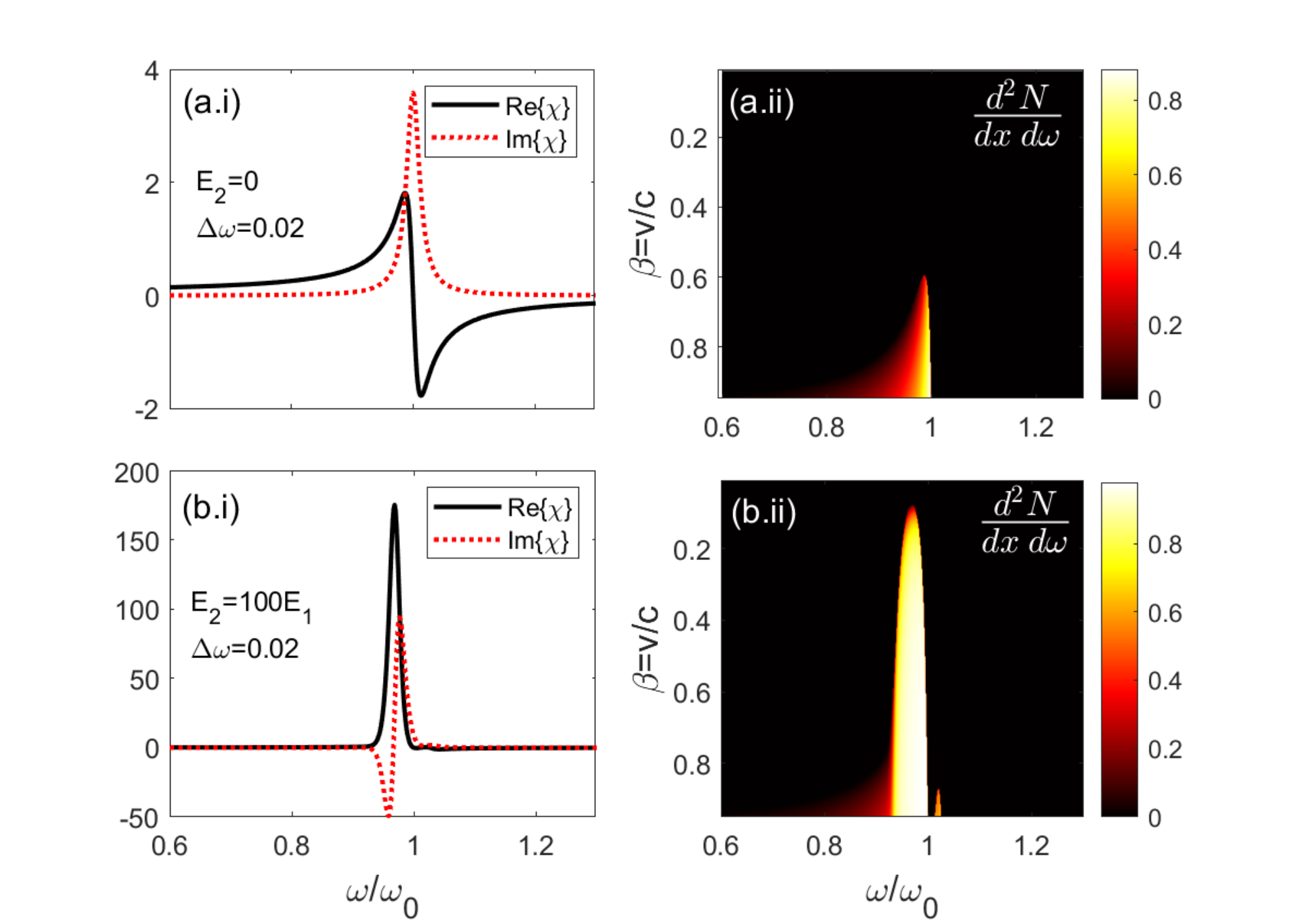}
\caption{(a.i) and (b.i) plot the dielectric susceptibility $\chi$ of the medium (polarization density) without and with the auxiliary Gaussian pulse of width $\Delta\omega=0.02\omega_0$, respectively. (b.i) susceptibility is largely enhanced compared to the bare susceptibility of the metal nanodimer (a.i). $\quad$ (a.ii) and (b.ii) plot the intensity of the Cherenkov radiation Eq.~(\ref{Np}) for charged particles moving at different speeds $\beta=v/c$. The electric field strength of the auxiliary pulse is assigned to be $E_2=100E_1$, with $E_1$ is the electric field of the Cherenkov emission. Enhanced susceptibility enables the radiation of particles moving at much smaller speeds. { We set $\phi=\pi/2$, in Eq.~(\ref{chi}), similar to Ref.~\cite{plasmon_index_enhancement}.}
\label{fig_w_beta_I_1}}
\end{center}
\end{figure}

\begin{figure}
\begin{center}
\includegraphics[trim=1.8cm 0 1cm 0 ,clip,width=0.5\textwidth]{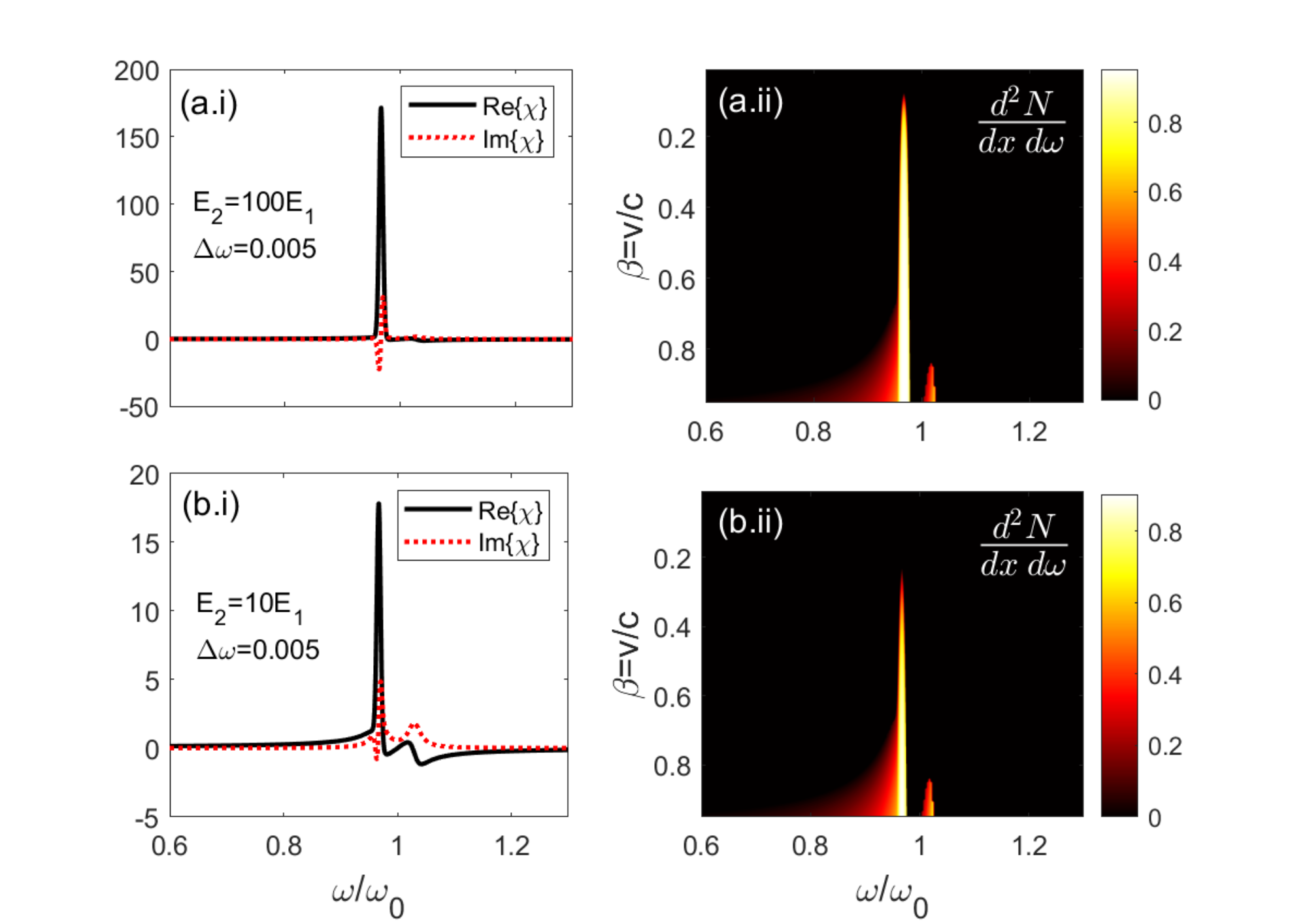}
\caption{ (a.i) and (b.i) Enhancement of the polarization density with a narrower $\Delta\omega=0.005\omega_0$ auxiliary pulse for two different (pump) electric field strengths, $E_2=100E_1$ and $E_2=10E_1$, respectively. (a.ii) and (b.ii) demonstrates the corresponding emission intensity. Radiation can be made in a (tunable) narrower spectrum via tuning the pump spectral width. Contribution from smaller particle velocities can also be tuned via the pump electric field $E_2$. { We set $\phi=\pi/2$, in Eq.~(\ref{chi}), similar to Ref.~\cite{plasmon_index_enhancement}}. \label{fig_w_beta_I_2}}
\end{center}
\end{figure}

At one point, we differentiate from the scheme of Ref.~\cite{plasmon_index_enhancement}, i.e. given in Fig.~\ref{fig_Lavrinenko}, a little. Because, the E-field induced via CR is not like probing the y-polarization oscillations (in the first nanorod) with a relatively small $E_1$ field. The CR is emitted spontaneously when the particle velocity along a direction exceeds the phase velocity of light. That is, we cannot simply consider sending a y-polarized probe pulse at frequency $\omega$. For this reason, we use an aux pump pulse of Gaussian shape in the frequency domain of spectral width $\Delta\omega=0.02\omega$, in Fig.~\ref{fig_w_beta_I_1}b, and $\Delta\omega=0.005\omega$ in Fig.~\ref{fig_w_beta_I_2}, i.e. $\tilde{F}_2=\sum_\omega F_2 e^{-i\phi} \exp[{-(\omega-\Omega)^2/(\Delta\omega)^2}] e^{-i\omega t}$. We determine the response $x_1$ by simply solving Eqs.~(\ref{x1})-(\ref{x2}) in the frequency domain $x_1(t)=\sum_{\omega} x_{1}(\omega) e^{-i\omega t}$~\cite{PS_frequencydomain}.

Fig.~\ref{fig_w_beta_I_1}(a.i) and \ref{fig_w_beta_I_1}(b.i) plot the real and imaginary parts of the susceptibility of a dielectric medium composed of such silver dimers with density $\rho \:\nu=0.02$, when the x-polarized pulse is off and on, respectively. The carrier frequency of the aux pulse, of width $\Delta\omega=0.02\omega_0$, is chosen to coincide with the frequency in Fig.~\ref{fig_Lavrinenko} where index enhancement with vanishing absorption appears { $\Omega=0.967\omega_0$}. In Fig.~\ref{fig_w_beta_I_2}(a.i), we use a sharper $\Delta\omega=0.005\omega_0$ aux pulse and we decrease (tune) the $E_2/E_1$ ratio in Fig.~\ref{fig_w_beta_I_2}(b.i). We observe that the index of such a medium can be ``continuously-tuned" order of magnitude at a narrow frequency range, if desired, at which one looks after observing tunable CR. Figs.~\ref{fig_w_beta_I_1}(a.ii),\ref{fig_w_beta_I_1}(b.ii),\ref{fig_w_beta_I_2}(a.ii) and \ref{fig_w_beta_I_2}(b.ii) are the CR intensities for particles moving at different speeds, possible via such a tuning. (The calculations are carried out in Sec.~\ref{sec:CR}.)

{ The ratio $E_2/E_1$ is a quantity that can be measured in an experiment. One can note that in an experiment, one does not need to tune the $E_2/E_1$ ratio. The $E_2/E_1$ ratio is, rather, needed for the analysis of the particles' velocity distribution, see Appendix~\ref{sec:AppE2E1} for some detailed discussion.} 

{ We remark that tuning the $E_2/E_1$ ratio between $E_2=100E_1$ and $E_2=10E_1$, in Figs.~\ref{fig_w_beta_I_2}(a.i) and \ref{fig_w_beta_I_2}(b.i), does not change the frequency where imaginary part of the dielectric function vanishes, i.e. $\omega=\Omega=0.967\omega_0$. This enables the continuous-tuning of the refractive index by circumventing the metallic losses causing the distortion in Cherenkov angle relation~\cite{GinisPRL2014Anisotropic,lin2018NaturePhys}.  }

In Fig.~\ref{fig_Lavrinenko}, one may appreciate that index enhancement is observable in a wider frequency range, of order $\gamma_{1,2}$. This allows one to choose a considerably wide range of frequencies for the index enhancement in processes where increased absorption is tolerable, i.e. when a smaller (partially absorbed) CR signal is sufficient where otherwise particle velocity is much below the cutoff.

\section{Control of the Cherenkov radiation} \label{sec:CR}

We calculate the total number of photons emitted via Cherenkov radiation, per unit traveling distance, { using the well-developed formula} 
\begin{equation}
\frac{dN_p}{dx}=\frac{1}{137c} \int_{\beta^2{\rm Re}[\epsilon(\omega)]>1} \: \omega\: d\omega\left( 1- \frac{{\rm Re}[\epsilon(\omega)]}{\beta^2|\epsilon(\omega)|^2}  \right), \label{Np}
\end{equation}
generalized to a dispersive medium~\cite{grichine2002energy,saffouri1984treatment,fermi1940ionization,sternheimer1953energy}. Here, the constraint $v/(c/n(\omega))>1$ is modified as $\beta^2{\rm Re}[\epsilon(\omega)]>1$~\cite{grichine2002energy,saffouri1984treatment} for an absorbing medium. { In Appendix~\ref{sec:Appendix-validity}, we present the refractive index for dielectric media composed of nanoparticles of sizes much smaller than the radiation wavelength. } { Such a material can be treated as an effective, homogeneous medium, so transition radiation can
be neglected~\cite{GinisPRL2014Anisotropic,duan2009research,galyamin2009reversed}. Because such media are shown to be homogenized to a very good approximation~\cite{cuabuzJOSAB2011homogenization,cuabuzPRL2007homogenization}. Central region of the medium, given in Fig.~\ref{fig_dimers}, can be unfilled to allow for unobstructed propagation of charged particles~\cite{GinisPRL2014Anisotropic}. }

In Figs.~\ref{fig_w_beta_I_1}(a.ii),\ref{fig_w_beta_I_1}(b.ii),\ref{fig_w_beta_I_2}(a.ii) and \ref{fig_w_beta_I_2}(b.ii), we plot the number of photons Cherenkov emitted at different frequencies by particles traveling at different speeds for the choice of a pump phase $\phi=\pi/2$~\cite{plasmon_index_enhancement}. 

When there is no aux field, particle velocities down to $v=$0.55$c$ is possible to emit a y-polarized CR~\footnote{Only a y-polarized CR emission occurs owing to the index enhancement. Because plasmons of the y-aligned nanorods, whose response to E-field is enhanced, couples to the y-polarized E-field much effectively~\cite{PS_anisotropy}.}, see Fig.~\ref{fig_w_beta_I_1}(a.ii), within the dimer density we choose.  When the aux field is turned on, by contrast, emission of a y-polarized CR from particles moving at speeds down to $v=0.1c$ becomes possible, see Fig.~\ref{fig_w_beta_I_1}(b.ii), via enhancement of the dielectric susceptibility $\chi=P/E$, see Fig.~\ref{fig_w_beta_I_1}(b.i). What more important is: the velocity threshold can be tuned \textit{continuously} by the intensity of the aux Gaussian pulse, e.g., compare Figs.~\ref{fig_w_beta_I_2}(a.ii) and \ref{fig_w_beta_I_2}(b.ii), after the detector is manufactured. The spectrum of the CR can also be tuned via tuning the spectral width and carrier frequency of the aux pulse even after the dimer ensemble is manufactured. { The frequency where absorption vanishes, i.e. $\omega=\Omega=0.967\omega_0$, does not change between $E_2=100E_1$ and $E_2=10E_1$ values. So, an aux pulse of carrier frequency  $\omega_c=\Omega=0.967\omega_0$  works in the absorption-free regime in the continuous tuning of the aux pulse ($\;\propto |E_2|^2\;$) intensity. }

\begin{figure}
\begin{center}
\includegraphics[scale=0.55]{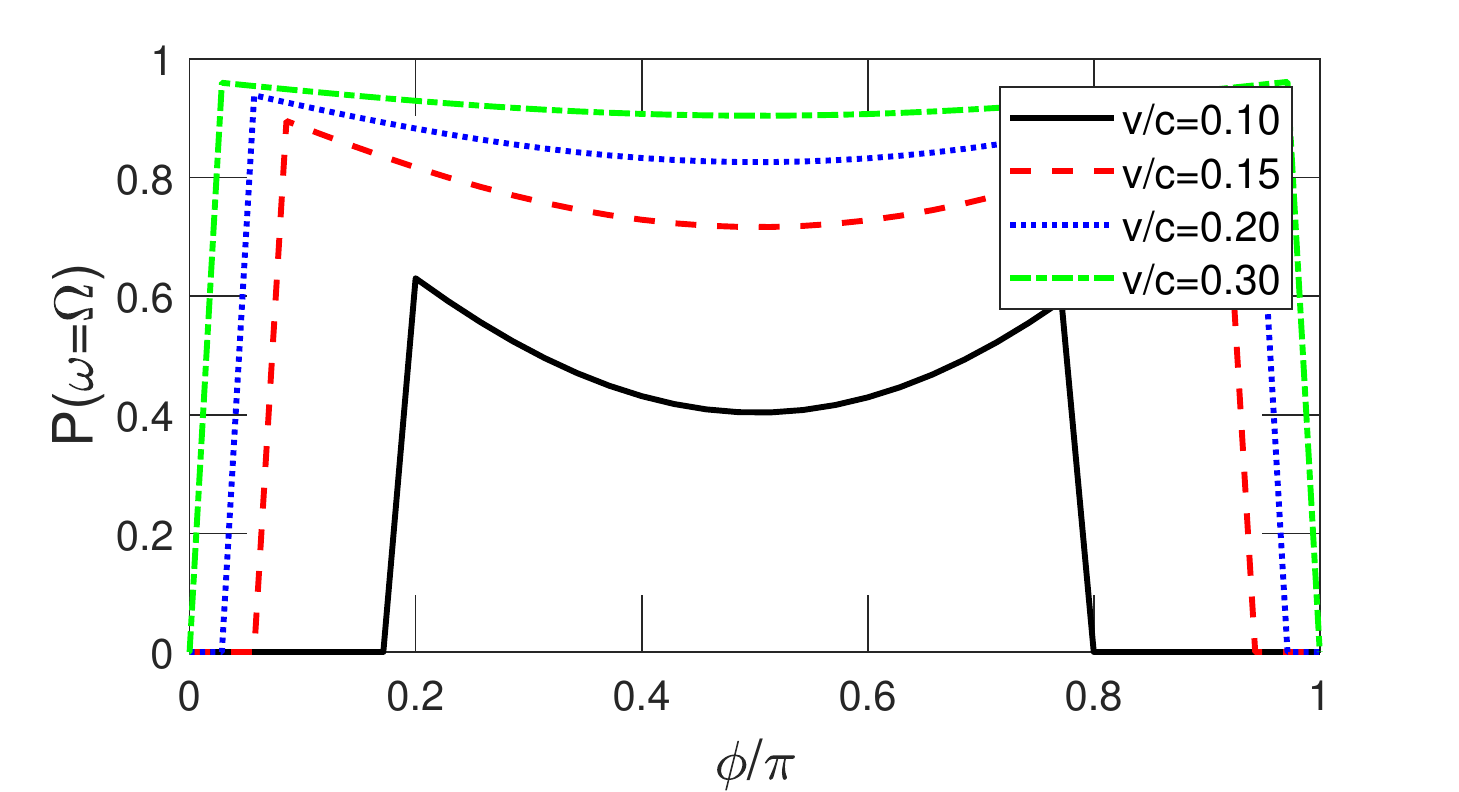} 
\caption{ Change of the emission intensity of charged particles at the enhancement frequency, { $\omega=\Omega=0.967\omega_0$}, for different phases $\phi$ of the auxiliary Gaussian pulse (or, equivalently, for different positions of the metal nanodimer). Different pump phases result in different refractive indices. { Aux pulse width is set to $\Delta\omega=0.005\omega_0$.}   \label{fig_phase_intens}}
\end{center}
\end{figure}

{ In Figs.~\ref{fig_Lavrinenko}-\ref{fig_w_beta_I_2}, we consider the phase $\phi=\pi/2$ for the aux pulse.} { The position of a dimer with respect to the aux pulse, however, alters the position-dependent phase at which the dimer operates. In Fig.~\ref{fig_phase_intens}, we plot the CR intensities for dimers operating at different pump phases $\phi$. When the periodicity of the dimers is not manufactured to match certain ratios of the aux pulse wavelength; a CR, possible to be emitted, is likely to feel a $\phi$-averaged index.} 

 { In Fig.~\ref{fig_Lavrinenko} we present the enhancement scheme for $\phi=\pi/2$~\cite{plasmon_index_enhancement} and consider the enhancement around { $\omega=\Omega\simeq0.967\omega_0$}. We remark that when $\phi$ assigns other values, the enhancement scheme, presented in Fig.~\ref{fig_Lavrinenko}, changes dramatically. At the frequency $\omega=\Omega\simeq 0.967\omega_0$, a value we choose referring to Fig.~\ref{fig_Lavrinenko}, may not correspond to an enhanced susceptibility. In Fig.~\ref{fig_phase_intens}, at $\phi/\pi=0.5$, i.e. $\phi=\pi/2$, all particle velocities $v=0.1\ldots 0.3$ generate CR. For a particle of velocity, e.g., $v=0.1c$, however, one cannot observe the CR in the phase range $\phi=0\ldots 0.18\pi$. Because in this phase range, according to Eq.~(\ref{chi}), the refractive index is not that high to suffice for creating a CR for a particle of velocity $v=0.1c$. For higher velocity values, e.g. $v=0.2c$, the refractive index, at { $\omega=\Omega\simeq 0.967\omega_0$}, becomes sufficient to generate CR emission in the phase range $\phi=0.04\ldots 0.96\pi$ via Eq.~(\ref{chi}) and (\ref{Np})  }. We also underline that the relative-phase, a single dimer operates, depends also on the $E_1$-field CR produces, i.e., not merely on the position of the dimer.

\begin{figure}
\begin{center}
\includegraphics[scale=0.55]{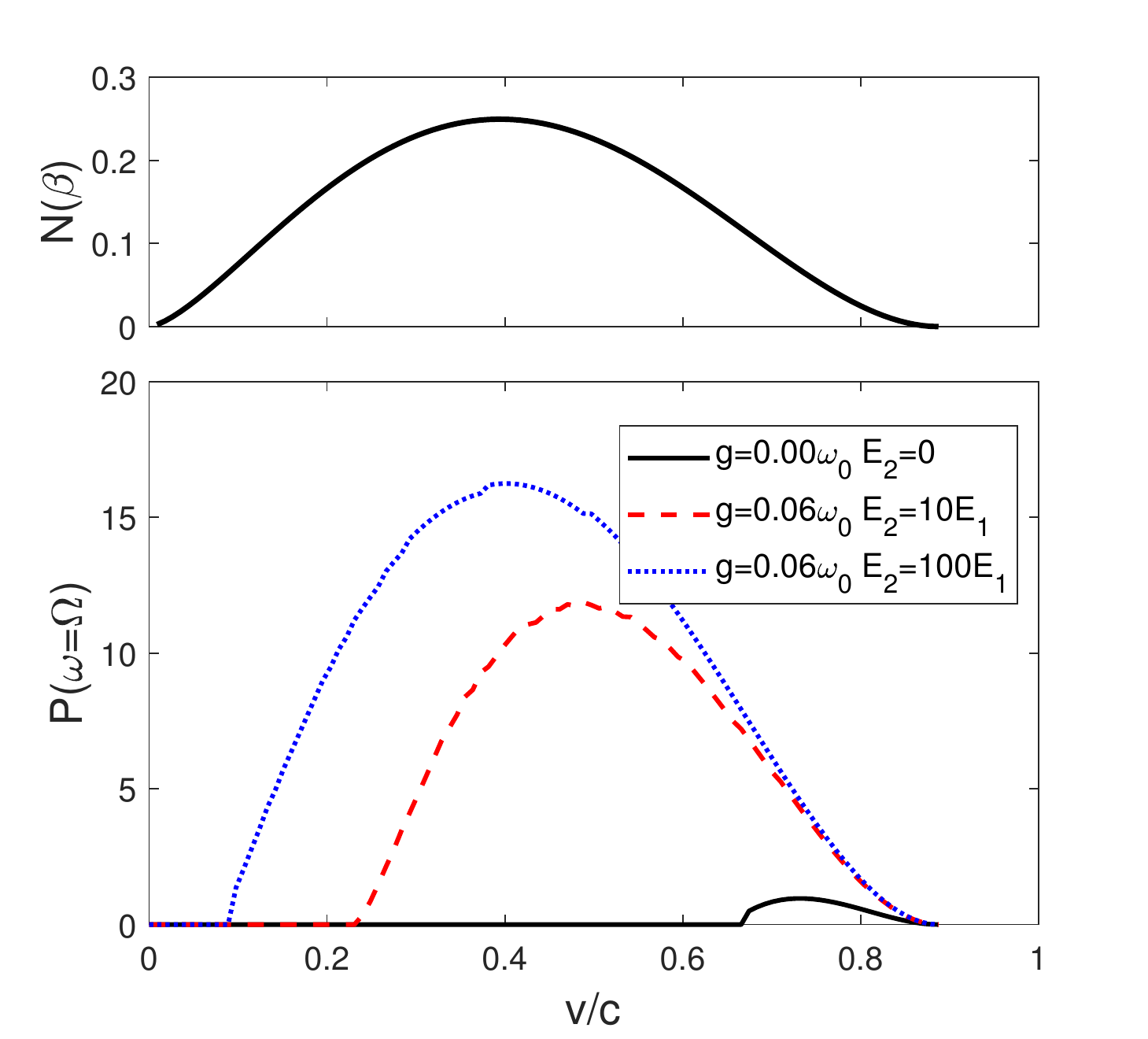} 
\caption{ (a) Velocity distribution of $^{18}$F emission, a material widely used in biological imaging~\cite{shaffer2017utilizing,spinelli2009cerenkov}. (b) Radiated power of the particles emitted by $^{18}$F, at the enhancement frequency { $\omega=\Omega=0.967\omega_0$}, for different field strengths $E_2$ of the auxiliary Gaussian pulse. Distribution depicted in (a) is used in the calculation and an average is considered over the phase $\phi$ of the auxiliary pulse (or, equivalently, over the positions of the nanodimers.) { Aux pulse width is set to $\Delta\omega=0.005\omega_0$.} \label{fig_beta_vs_Pw}}
\end{center}
\end{figure}

In Fig.~\ref{fig_beta_vs_Pw}, we plot the CR emission at different particle velocities by considering the energy distribution of a ${}^{18}F$~\cite{levin1999calculation} an isotope widely used, e.g., in imaging applications~\cite{shaffer2017utilizing,spinelli2009cerenkov}. In Fig.~\ref{fig_beta_vs_Pw}b, we also perform an average over the phase $\phi$ considering a (general) periodic structure whose periodicity does not match the aux pulse wavelength. We also note that dimers ($\sim$30 nm) are much smaller compared to the CR emission, at optical wavelengths, e.g., $\sim$500 nm, which overlaps many of such dimers in 3D.

\begin{figure}
\begin{center}
\includegraphics[scale=0.55]{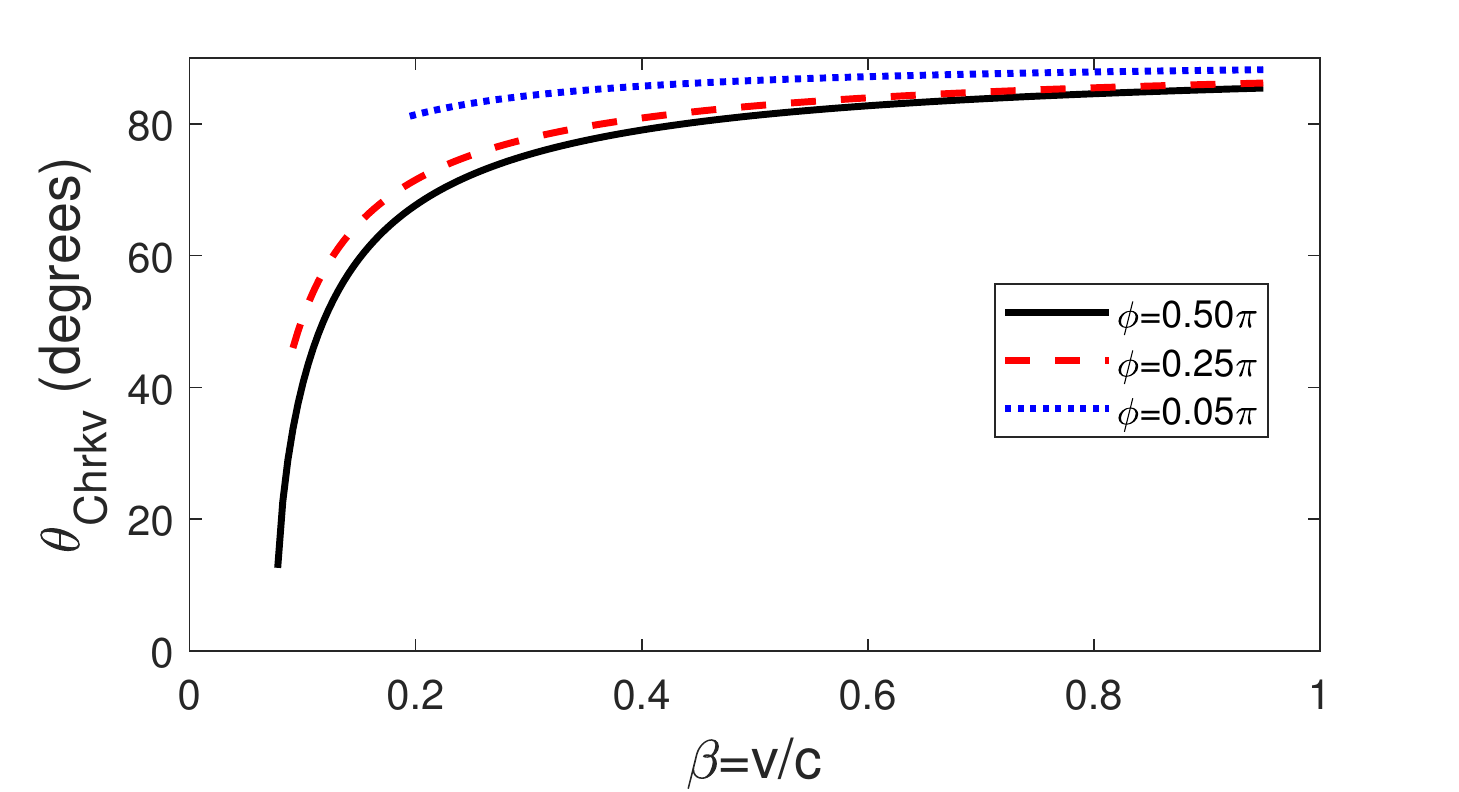} 
\caption{Direction of the Cherenkov radiation $\theta_{\rm \scriptscriptstyle Chrkv}$ with respect to the velocity of the charged particle for different values of the phase $\phi$ of the Gaussian pulse. { Carrier frequency and the spectral width of the aux pulse are set to $\omega_c=\Omega=0.0967\omega_0$ and $\Delta\omega=0.005\omega_0$, respectively.  } \label{fig_beta_vs_thetaC}}
\end{center}
\end{figure}

In Fig.~\ref{fig_beta_vs_thetaC}, we plot the CR angle $\cos\theta_{\rm \scriptscriptstyle CR}=\sqrt{{\rm Re}[\epsilon(\omega)]}/\beta |\epsilon(\omega)|$ in a dispersive medium~\cite{grichine2002energy,saffouri1984treatment} emitted from a charged particle moving at different velocities. We observe that $\theta_{\rm \scriptscriptstyle CR}$ from slower-moving particles depend on the pump-phase dramatically, { comparing the results for $\phi=0.50\pi$ with $\phi=0.05\pi$~\footnote{ For $\phi=0$, the condition for the Cherenkov radiation is not satisfied for any $\beta$ value.}}. { Therefore, for periodic structures, where even a wavelength contains hundreds of nanodimers, one needs to consider an average over the phase (dimer positions) $\phi$, before calculating the CR angle relations. This may require calibration with a known source before carrying out the actual experiment. } Therefore, beyond observing CR from slow-moving particles and CR intensity enhancement, $\theta_{\rm \scriptscriptstyle CR}$ analysis is better to be held with periodic structures where a $\phi$-average can be considerable for optical CR. 

{ {\it Measurement scheme}--- We note that the propagation of the aux-pulse-controlled CR emission, which is y-polarized, can be only in the x-z plane. Since the incident particles are propagating along the x-direction, the controllable (continuously-tunable) CR emission is tilted along the z-direction via a CR emission angle $\theta_{\rm \scriptscriptstyle CR}$. In order to observe the distribution of the CR angle, one can filter-in (collect) the y-polarized CR emission. At this stage, there appears a complicating effect: high-velocity particles can also emit y-polarized CR whose origin is not related with the y-polarized nanorod holding a controllable index, e.g. due to the background solution itself. This complication (actually any such effect leading to aux pulse-invariant CR emission) can be traced out from the CR via the index control phenomenon as follows. Cherenkov radiation angle measurement can be carried out (calibrated) for a turned off auxiliary (y-polarized index-controlling) pulse and all other particle spectroscopies can be carried out referring this calibration point. }

\section{Summary} \label{sec:summary}

In summary, we show that index enhancement in plasmonic metamaterials~\cite{plasmon_index_enhancement} can be utilized to gain control over Cherenkov radiation. An auxiliary x-polarized Gaussian pulse controls the refractive index of the medium for a y-polarized Cherenkov emission. Beyond enhancing the overall Cherenkov radiation and enabling radiation from much slower particles; { such a setup enables spectral analysis of the particle velocities as if there exist many different particle detectors. Because, different intensity values of the auxiliary pulse operates different particle detector media and this allows the data correction~\cite{DataCorrectionDetector}.}
{ Moreover, vanishing absorption heals the distorting effects on the Cherenkov angle relation which are caused by the absorption of metal nanostructures. }

In addition, the frequency width of the Cherenkov radiation can also be tuned with the width of the auxiliary pulse. The utility of the technique, we introduce here, can be well appreciated by considering that: the phenomenon of Cherenkov radiation has implementations in particle detection and medical imaging.

\begin{acknowledgments}
Contribution of the authors MG, YLC and MET are 25\%, 25\% and 50\%, respectively. We thank Mehmet Y{\i}ld{\i}z for instructive discussions. YLC would like to acknowledge the support of Ministry of Science and Technology (MOST) of Taiwan. MET gratefully thank Atac Imamoglu, Manuela Weber-Semler and Mehmet Erbudak for their hospitality. MET and MG acknowledge support from TUBITAK 1001 No:117F118.  MET acknowledges support from TUBA GEBIP 2017. 
\end{acknowledgments}

\appendix

{

\section{The $E_2/E_1$ ratio} \label{sec:AppE2E1} 

There are two ways to perform an $E_2/E_1$ arrangement.

Before using the manufactured medium with the auxiliary pulse, one can learn the properties of the medium without the aux pulse. One needs to perform this only once after manufacturing the detector. One can use an already known source for the detector calibration and obtain the curve in Fig.~\ref{fig_w_beta_I_1}(a.ii), e.g., by recording at a specific wavelength. Then, in the same calibration process, one can turn on the aux Gaussian pulse, for a trial $E_2$, and records the new $E_1$. This way, one can gain an initial information on the CR response/characteristics of the device before using it with unknown velocity particles.

Actually, tuning (knowing) $E_2/E_1$ before an experiment is not necessary. Because the purpose of this new device is to provide a continuous spectral analysis on the velocity distribution of the particles, e.g., with known (measured) $E_2/E_1$ values. $E_2$ and $E_1$ ($I_2$ and $I_1$) are measured in an experiment with an unknown radioactive source. Hence, one already records the continuous data (parameter) of $E_2/E_1$. One can also compare them with the data already obtained from the calibration with the known source, even though this is not necessary. Besides the continuous spectral analysis of the unknown source ---which is the true aim of the proposed device--- if one aims to tune the $E_2/E_1$ ratio, e.g. wants to operate the detector at a specific regime, the experimentalist can actually do this via a continuous measurement of the $E_2/E_1$ ratio, e.g., at a specific wavelength.

\section{Effective Polarizibility and Cherenkov Radiation} \label{sec:Appendix-validity}

Cherenkov radiation in particle detectors, employing nanoparticles, is a well-developed formalism which has been studied for a decade. The effective medium/polarization methods, which rely on calculating the average electric/polarization fields~\cite{Silveirinha2012cherenkov,vorobev2012nondivergent,silveirinha2006nonlocal,silveirinha2005homogenization,so2010cerenkov,silveirinha2017metamaterials, morgado2015analytical,tyukhtin2011effective,tyukhtin2014radiation,nefedov2005propagating,GinisPRL2014Anisotropic}, have already been demonstrated to work well in nanowire/nanoparticle arrays in such parameter regimes where the size of the nanostructures are much smaller than the CR wavelength. 

These methods are able to treat also the periodicity of the wires, e.g. by inserting the Floquet mode solutions (due to periodicity) into polarization/electric field averages. While many recent works concentrate on effective permittivity of periodic nanoparticle structures, owing to the unique optical features of such metamaterials, Ref.~\cite{yurtsever2008formation} presents a clear demonstration of the effective polarization for the arbitrary distribution of, e.g. silicon, nanoparticles in a medium where the theoretical treatment fits the experimental data. Thus, although we consider a periodic structure of the silver dimers here, arbitrarily distributions of such dimers, e.g. in solutions, can also be utilized as index-controlled continuously-tunable particle detectors. Dimers can be aligned using magnetized metallic material~\cite{anderson2019magnetic,Enzo2017giantMagnetic}.


We calculate the average polarization of the medium, e.g. in a wavelength size, which contains many of such nanoparticles. The maximum length of a dimer is $L$=30 nm, which is much smaller than an optical wavelength. The mean separation between the centers of the dimers is crudely $\sqrt[3]{50}\times L\simeq$110 nm, since a unit volume contains $\rho \nu$=0.02 of such dimers. We note that this, i.e. $\rho \nu$=0.02, is an arbitrary value we assign in the manuscript, where a denser/lighter ensemble would be possible. The volume of a dimer is approximately 30$\times$30$\times$10=8100 ${\rm nm}^3$. The volume of a typical wavelength is ${\rm V}_{\lambda}$=${\rm 500}^3 {\rm nm}^3$. Hence, in a single wavelength volume there exists 310 of such dimers. This value can be increased/decreased via other choices for the density of dimers. That is, for 310 of such dimers in a wavelength, assigning an effective polarization (even for arbitrarily-ordered dimers) is rather like defining a polarization density in the books of electromagnetic theory~\cite{griffiths2005introduction,jackson2007classical}. 

Actually, for the validity of the simple treatments of Cherenkov radiation in such media, the critical issue is a concept related with the ``length of formation"~\cite{yurtsever2008formation,zolotorev2000classical,klein1999suppression}. A Cherenkov wave takes on the character of a radiation, that is no longer tied to its source, after the particle travels the formation length~\cite{yurtsever2008formation,zolotorev2000classical,klein1999suppression} which is order of a wavelength or less. That is, the condition for CR is (i) material to be extended for some degree (i.e. formation length) for the CR to be formed, and (ii) in the packed nanoparticle system it must be provided that nanoparticle dimensions are much shorter than the propagation wavelength. Satisfying these conditions, a packed nanoparticle system can be treated as an effective medium in the context of formation [13]. Thus, for CR to emit, the medium has to extend at least a formation length; i.e. a single nanoparticle alone smaller than the formation length cannot emit the CR and a continuation of such particles is necessary. After this condition satisfied, an effective medium treatment regarding the wavelength of the virtual photons to be emitted, e.g. 310 dimers in a wavelength volume, can be considered.

On top of all these discussions, there will certainly appear complicating effects. These, however, are certainly out of the scope of this work which aims to present the audience a basic (as the first) demonstration of the implementation of the index-enhancement phenomenon to particle detectors allowing continuously-tunable spectral analysis ---not targeting a rigorous simulation, revealing all effects, in an already conducted experiment. 

}

\bibliography{bibliography}

\begin{thebibliography}{75}
\bibitem [{PS_({\natexlab{a}})}]{PS_frequencydomain}%
  \BibitemOpen
  \href@noop {} {}\bibinfo {note} {We determine the coefficients $x_i(\omega)$
  by using the expansion $\sum_\omega x_i(\omega)$ in Eqs.~(\ref{x1},\ref{x2}).
  We discretize the 1D frequency space and solve such linear
  equations.}\BibitemShut {Stop}%
\bibitem [{PS_({\natexlab{b}})}]{PS_anisotropy}%
  \BibitemOpen
  \href@noop {} {}\bibinfo {note} {In an extended study on the further
  implementations of the index enhancement phenomenon on the behavior of the CR
  emission, one may consider the effect of the dielectric anisotrop which is
  demonstrated to cast a threshold-less CR.}\BibitemShut {Stop}%
\end{thebibliography}

\end{document}